\documentclass[aps,
		pra,
		reprint,
		twocolumn,
		superscriptaddress,
		shortbibliography,
		nofootinbib,
		floatfix,
		notitlepage
		]{revtex4-2}
\pdfoutput=1

\usepackage{graphicx}
\usepackage[colorlinks=true, urlcolor=blue, citecolor=blue, linkcolor=blue]{hyperref}
\usepackage{color}
\usepackage{bm} 
\usepackage{newtxtext, newtxmath}
\usepackage{tabularx}
\usepackage{geometry}		
\usepackage{setspace}
\usepackage{cleveref}
\usepackage{textcomp}
\newcommand{\be}{\begin{equation}}
\newcommand{\ee}{\end{equation}}
\newcommand{\bea}{\begin{eqnarray}}
\newcommand{\eea}{\end{eqnarray}}
\newcommand{\bi}{\begin{itemize}}
\newcommand{\ei}{\end{itemize}}

\hypersetup{colorlinks=true,citecolor=blue,urlcolor=magenta}

\newcommand{\vkap}{\varkappa}

\newcommand{\trm}[1]{\textrm{#1}}
\newcommand{\tsf}[1]{\textsf{#1}}
\newcommand{\av}[1]{\langle{#1}\rangle}

\newcommand{\Ecr}{F_{\tiny\textsf{qed}}}

\newcommand{\figref}[1]{Fig.\,\ref{#1}}
\newcommand{\figrefa}[1]{Fig.\,\ref{#1}\,a)}
\newcommand{\figrefb}[1]{Fig.\,\ref{#1}\,b)}
\newcommand{\eqnref}[1]{Eq. (\ref{#1})}

\newcommand{\nn}{\nonumber}

\begin{document}

\title{Revealing signals of higher-order nonlinear showers in particle-laser collisions}

\author{T. G. Blackburn}
\email{tom.blackburn@physics.gu.se}
\affiliation{Department of Physics, University of Gothenburg, SE-41296 Gothenburg, Sweden}
\author{B. King}
\email{b.king@plymouth.ac.uk}
\affiliation{Centre for Mathematical Sciences, University of Plymouth, Plymouth, PL4 8AA, United Kingdom}
\author{M. Samuelsson}
\email{mathias.samuelsson@physics.gu.se}
\affiliation{Department of Physics, University of Gothenburg, SE-41296 Gothenburg, Sweden}

\date{\today}

\begin{abstract}
Several high power laser facilities are reaching field strengths where leading order strong-field quantum electrodynamical (QED) processes can be measured in the non-perturbative regime for the first time. At very high, as yet unobtainable in the laboratory, field strengths, the contribution of higher-order processes is predicted to dominate,  implying a breakdown of current calculational methods. Focusing on nonlinear showers and considering currently available experimental parameters, we find that if the momentum spectrum of the \emph{incident} particles is well known, asymmetries in the \emph{outgoing} particle spectrum  may provide a useful signature of higher orders of nonlinear phototrident, trident and Compton scattering. These signatures could be used by experiment to test how accurate the current calculational framework is when applied to strong-field QED at higher orders.
\end{abstract}

\maketitle

\vspace{-0.9cm}
\section{Introduction}
Shortly after invention of the laser, calculations were made that showed nonlinear Compton scattering \cite{Nikishov:1964zza,Brown:1964zzb} and nonlinear Breit-Wheeler pair-creation \cite{1962JMP.....3...59R} could be accessed in the laboratory using sufficiently strong electromagnetic fields. As laser technology develops, new records continue to be set for the highest attainable electromagnetic field intensity in the laboratory \cite{2021Optic...8..630Y}. Coupled with laser-plasma accelerated electron beams, `all optical' experiments are starting to be able to probe the regime of strong-field QED \cite{cole.prx.2018,poder.prx.2018,CoReLs24,Los:2024ysw}. Whilst such experiments can reach high laser intensities, they still face significant challenges such as reproducibility of results (due to beam jitter), electron beam quality, and the limits of data accumulation due to the low repetition rate of the lasers. However, if these challenges can be overcome, the precision of all-optical experiments and hence strong-field QED science reach can be significantly improved. Indeed the first observation of the nonlinear trident process in the multiphoton regime, well-approximated by nonlinear Compton scattering followed by nonlinear Breit-Wheeler pair-creation creation, was in the E144 experiment \cite{bula.prl.1996,burke.prl.1997,Bamber:1999zt}, which combined a stable and well-characterised $46.6\,\trm{GeV}$ electron beam accelerated by the conventional method of radio frequency cavities and an optical laser pulse with intensity parameter $a_{0}\approx 0.3$. Natural successors to this experiment include E320 at FACET-II \cite{Salgado:2021fgt}, which is aiming to reproduce these results in the regime $a_{0} \gtrsim 1$, where all-orders of interaction in the charge-field coupling, $a_{0}$, must be taken into account. Another planned experiment is LUXE at DESY \cite{abramowicz.epjst.2021,abramowicz.2023}, which would have a dedicated set-up to accumulate large amounts of data and allow for more detailed mapping of the perturbative to non-perturbative transition and higher-order processes. 

In the current paper, we use the Ptarmigan simulation code~\cite{blackburn.pop.2023} and direct calculation to demonstrate how signals of higher-order strong-field QED processes may be generated using electron and photon sources of comparable energies to those that will be available in all-optical experiments. In Sec. II, we give an overview of the ordering of how the hierarchy of higher-order processes in an intense electromagnetic background is arranged, and in Sec. III we focus on the case study of nonlinear phototrident ($\gamma\to e^{+}e^{-}+\gamma'$). Using skewness in the electron-positron lightfront momentum distribution as a guiding signal, we investigate requirements on the bandwidth of photon sources necessary to observe the presence of higher orders. In Sec. IV some example observables for measuring nonlinear Compton scattering ($e \to e \gamma$) and nonlinear trident ($e \to e^+ e^- e$) at higher order (i.e. at least third order in dressed vertices) are also investigated. In Sec. V we end with the conclusion that increasing  overall accuracy in all-optical experiments can allow us to test theory in this transition regime between single dressed vertex processes and the fully non-perturbative QED regime predicted by the Ritus-Narozhny conjecture \cite{Fedotov:2016afw,Podszus:2018hnz,Ilderton:2019kqp,Mironov:2020gbi,Mironov:2021ohk,Heinzl:2021mji,Torgrimsson:2021wcj}, thereby building upon recent science cases made in the literature \cite{weber.mre.2017,gales.rpp.2018,mp3,turner.2022,LUXE:2025wuo,Sarri:2025qng}.

\section{Overview}

\begin{figure}[h!!]
\includegraphics[width=8cm,draft=false]{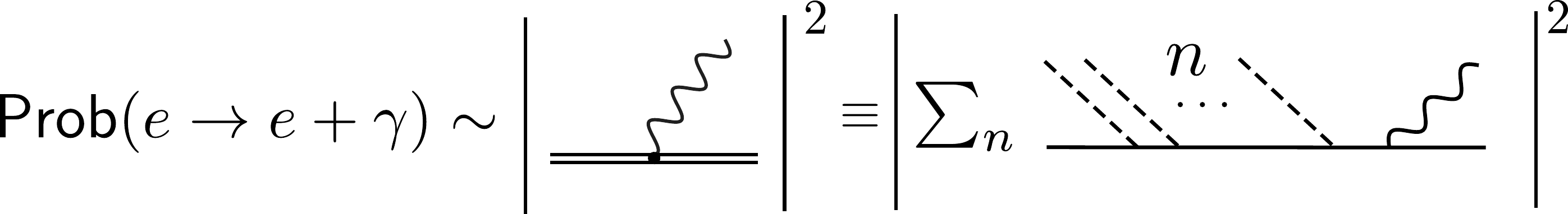}
    \caption{`Nonlinear' Compton scattering includes all orders of interaction between the laser field (dashed lines) and the electron/positron.}
    \label{fig:NLCdiag1}
\end{figure}

The `strong-field' regime of QED is characterised by the strong-field parameter $\chi$ reaching $\chi \sim O(1)$ (see reviews e.g. \cite{dipiazza.rmp.2012,narozhny15,gonoskov.rmp.2022,fedotov.pr.2023}). For an electron beam colliding with an intense laser pulse $\chi \approx F_{\tsf{rf}}/\Ecr$, where $F_{\tsf{rf}}$ is the field strength in the rest frame of an electron and $\Ecr = m^{2}c^{3}/e\hbar \approx 1.6\times 10^{16}\,\trm{Vcm}^{-1}$ is the Schwinger critical field strength. (In a plane wave the relation between $\chi$ and $F_{\tsf{rf}}$ becomes exact.) When $\chi \sim O(1)$, it becomes probable that electron-positron pairs can be produced by photons via the nonlinear Breit-Wheeler process, pairs can be produced by an electron via the nonlinear trident process, and recoil on electrons and positrons from Compton scattering makes a significant difference to emitted photon energies. In a high-energy electron-laser collision, the strong-field parameter is approximately given by a product of the (laser) intensity parameter, $a_{0}$ and the energy parameter $\eta$:
\[
\chi = a_{0} \eta; \qquad \eta = \frac{\gamma\omega}{m} (1+\cos\theta),
\]
where $\gamma$ is the Lorentz gamma factor, $\omega$ is the laser frequency and $\theta$ is the collision angle between the electron and laser beams ($\theta=0$ for a `head-on' collision). With a motivation from all-optical experiments, we will assume a characteristic particle energy of around $8\,\trm{GeV}$~\cite{gonsalves.prl.2019} and a laser wavelength of $\lambda= 0.8\,${\textmu}m, giving $\eta \sim 0.1$ and $\omega = 1.55\,\trm{eV}$.
The effective coupling between the electron and the laser field is $a_{0} = e (F/m\omega)$ where $F$ is the laser field strength in the laboratory frame and $e$ and $m$ are the charge and mass of a positron. If $a_{0} \ll 1$, only the leading-order interaction with the laser field needs to be taken into account, but if $a_{0} \not \ll 1$ all orders of interaction with the laser field must be included, as depicted in \figref{fig:NLCdiag1}. This is sometimes referred to as `\emph{non-perturbativity at small coupling}' in analogy with gluon saturation in QCD \cite{FCC:2018byv}, and is represented as a single dressed vertex.

Another type of non-perturbativity is hypothesised to arise when the probability for nonlinear QED processes themselves becomes large enough that the series in \emph{dressed} vertices can no longer be truncated. It was conjectured by Ritus and Narozhny, that such a parameter regime may arise when $\chi \sim 1/\alpha^{3/2} \sim O(1000)$, at which point the Furry expansion method is expected to break down \cite{Fedotov:2016afw}.

\begin{figure}[h!!]
\centering
\includegraphics[width=8.5cm,draft=false]{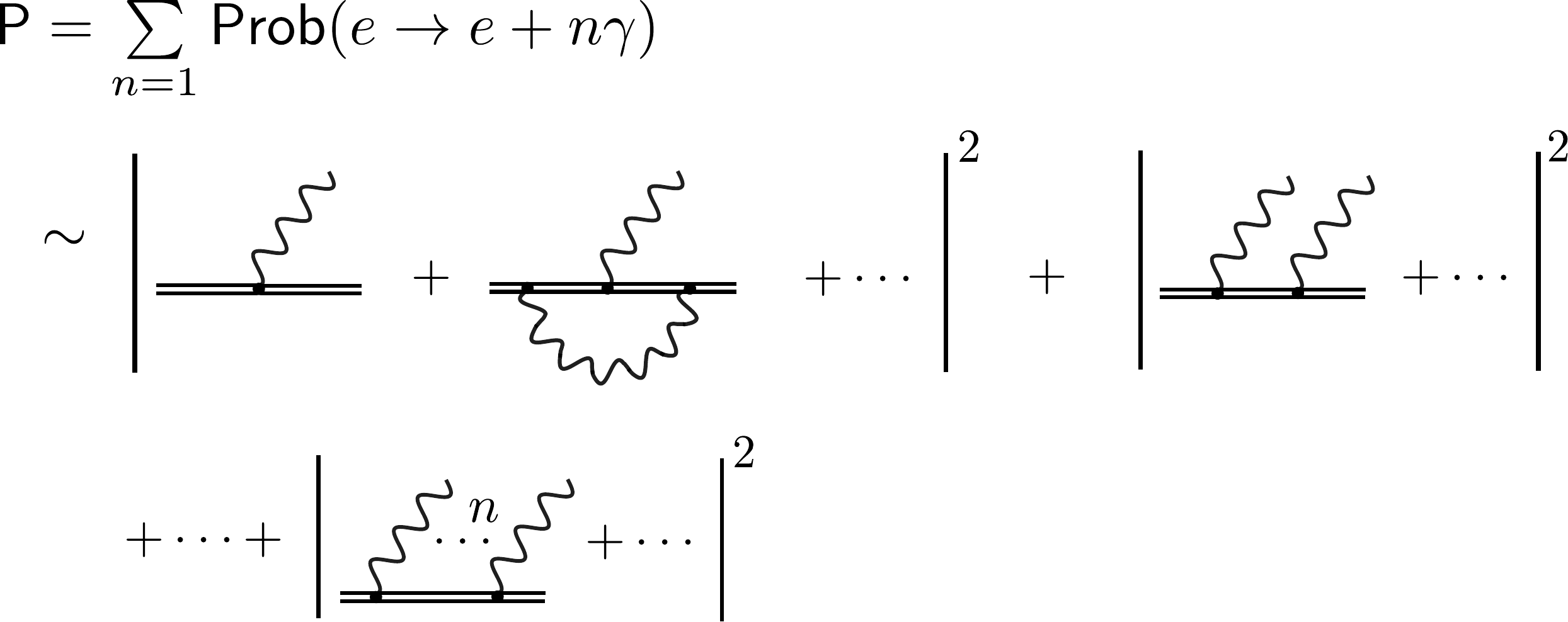}
\caption{Nonlinear Compton Shower. Diagrams in ellipsis contain contributions that are higher orders in $\alpha$ (i.e. contain loops).
}\label{fig:loops1}
\end{figure}

In the current paper, we are interested in the transition regime between these two cases, in which higher order strong-field QED processes are probable enough that multiple dressed orders must be taken into account. We focus on nonlinear showers, where the incident and any produced charges decelerate when scattering in the laser field and radiate photons and electron-positron pairs. (In contrast to nonlinear avalanches where charges may be significantly reaccelerated by the laser field \cite{King:2013zw,Mironov:2014xba,Mironov:2025nhk}.) The processes involved in calculating the total probability of a nonlinear shower dominated by nonlinear Compton scattering (NLC) are sketched in \figref{fig:loops1}.

\begin{figure}[h!!]
\centering
 \includegraphics[width=8.5cm,draft=false]{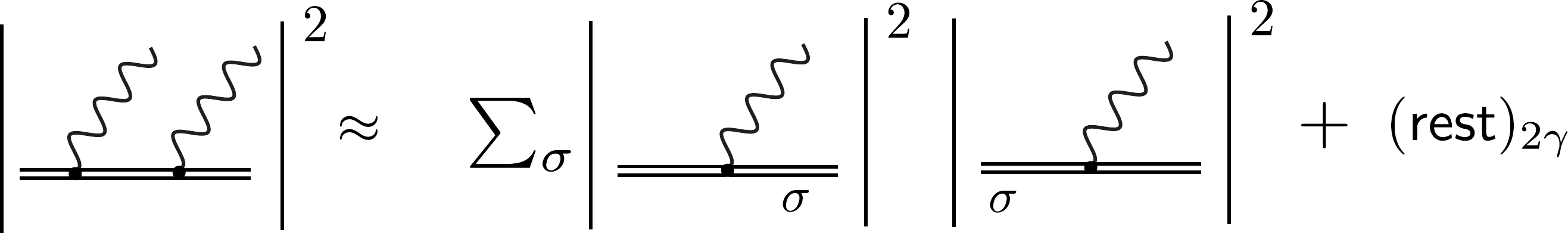}
\caption{
 Representation of the factorisation of double nonlinear Compton scattering in sequential first-order nonlinear scattering which includes a sum over the spin state $\sigma$ of the propagating particle. The `rest' includes non-propagating contributions and interference terms.
}\label{fig:factorDiag1}
\end{figure}

We will calculate these processes mainly by using the single-particle code Ptarmigan \cite{Blackburn:2023mlo}. This employs standard methods of propagating particles on classical trajectories whilst Monte-Carlo sampling at each timestep rates for single dressed-vertex QED processes, which if triggered, correct the classical dynamics~\cite{ridgers.jcp.2014,gonoskov.pre.2015}. This effectively corresponds to factorising higher-order processes in terms of chains of first-order processes joined by a particle propagating on-shell. Justification for this factorisation into an $n$-step process arises 
from direct studies of second-order processes \cite{Hu:2010ye,Seipt:2012tn,Mackenroth:2012rb,King:2013osa,Dinu:2017uoj,Mackenroth:2018smh,King:2018ibi,Dinu:2019wdw,Torgrimsson:2020mto,Kaminski:2022uoi} i.e. $n=2$. It is found that at large intensity parameter $a_{0}\gg 1$, the total probability for photon emission, $\tsf{P}$, can be expanded in the parameter $1/a_{0}$, with the leading-order term (in the Laurent series) given by $\sim a_{0}^{2}$ being proportional to the $n$-step part. (This factorisation has been implemented to calculate higher order processes and benchmarked \cite{Torgrimsson:2020gws,Torgrimsson:2021wcj}.) This dominant part is included in modern simulation frameworks; the remainder terms are neglected, indicated by dropping the `rest' term in \figref{fig:factorDiag1}.

\begin{figure}[h!!]
\vspace{-0.5cm}
\[
\tsf{P} \sim \sum_{k=1} \tsf{P}^{[k]}; \qquad \tsf{P}^{[k]} = (\alpha \chi \Phi)^{k}\mathcal{I}^{[k]}
\]
\centering
\includegraphics[width=8.5cm,draft=false]{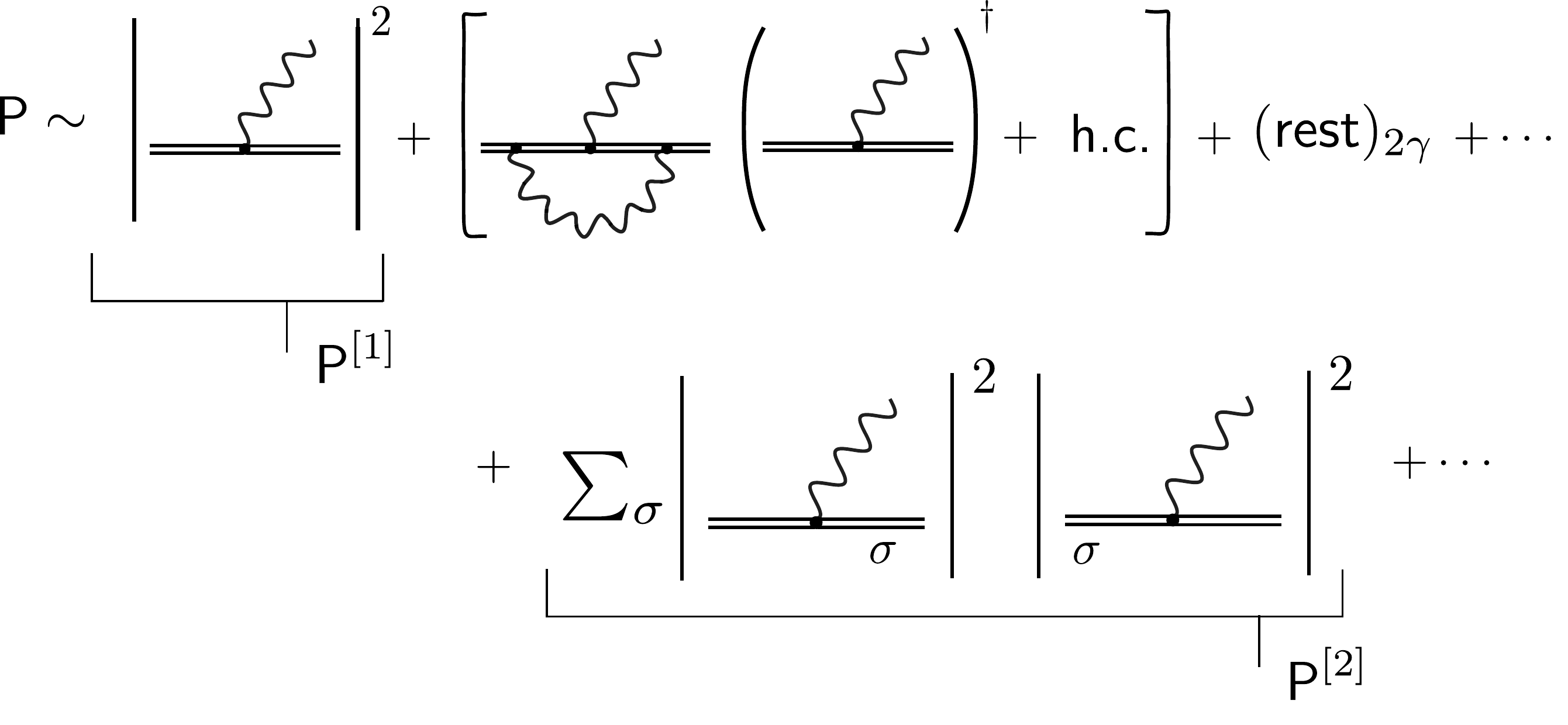}
\caption{
 The total probability for nonlinear Compton shower expanded as a series in pulse duration $\Phi$. At each order, the dominant contribution when $a_{0}\gg1$ is given by $\tsf{P}^{[k]}$ i.e. the sequential channel from $n$-fold iterations of single nonlinear Compton scattering joined by a real propagating particle. Numerical simulation includes the dominant contribution but does not in general explicitly include loop interference or off-shell tree-level contributions.
}\label{fig:SeriesDiag1}
\end{figure}

For the case of the nonlinear Compton shower in \figref{fig:loops1}, we can reason about the scaling of higher-order processes in the following way. Suppose $a_{0} \gg 1$ so that the probability is well-approximated by an integration over the NLC rate in a locally constant crossed field and the formation length is much shorter than the laser wavelength. If $\chi \ll 1$, the NLC rate $\propto \alpha\chi$ (this holds quite well even up to $\chi =1$ \cite{Seipt:2020diz}). If the laser pulse is long enough to have $n$ NLC events from a single electron, then the scaling is $\sim \Phi^{n}$, where $\Phi$ is the phase duration of the pulse. So in this regime, $n$-fold nonlinear NLC scales as $\sim (\alpha \chi \Phi)^{n}$. Although these arguments are rudimentary, 
they emphasise that if the pulse is long enough, or the field strength high enough, higher-order dressed processes dominate (see \figref{fig:SeriesDiag1}).

A consistent calculation of higher-order processes requires the inclusion of dressed loops and resummation of infinite series of diagrams. This is a challenging area of theory that requires experimental input and motivates our study in the current paper.

In the following, we focus our analysis on demonstrating signals for nonlinear photo-trident and then give some examples of signals of higher order nonlinear Compton scattering and nonlinear trident.

\section{Nonlinear Phototrident}
\begin{figure}[h!!]
\includegraphics[width=7cm,draft=false]{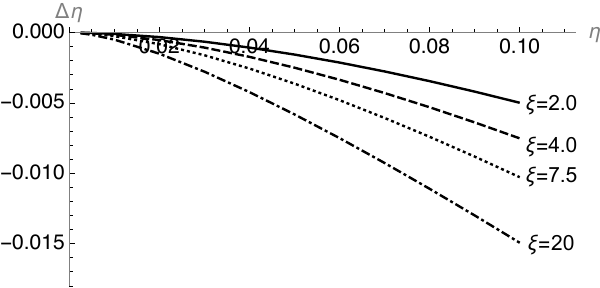}
    \caption{The average change in energy parameter, $\Delta \eta$, of an electron or positron with initial energy parameter, $\eta$, when it nonlinear Compton scatters in an eight-cycle sine-squared plane wave pulse. (Calculated using the locally constant field approximation, LCFA.)}
    \label{fig:NLC-momentum1}
\end{figure}

In photon-laser collisions, the leading-order (LO) process for creating pairs is nonlinear Breit-Wheeler (NBW) $\gamma \to e^{+}e^{-}$. If many photons of similar energy parameter create electrons and positrons via the LO process, the electron and positron lightfront momentum spectrum will be symmetric and peaked around half of the photon lightfront momentum. Defining $s=\vkap \cdot q/\vkap \cdot \ell$ as the lightfront fraction where $q$ and $\ell$ are the positron and photon momenta ($\omega = \ell^0$) and $\vkap$ is the laser wavevector, the lightfront momentum spectrum of the positron (electron) is peaked around $s=1/2~$ (${1-s=1/2}$). There is a symmetry $s \to 1-s$ in the positron and electron spectrum which follows from the charge-parity symmetry of QED. The next-to-leading-order (NLO) process of creating pairs in photon-laser collisions is nonlinear Breit-Wheeler followed by a photon emission from the electron or positron, sometimes referred to as (lowest order) `nonlinear phototrident' \cite{MorozovNarozhnyiPhTr,Torgrimsson:2020mto}: $\gamma\to e^{+}e^{-}+\gamma'$. This NLO process no longer displays the symmetry in the positron lightfront momentum around $s=1/2$; the momentum is not altered by a nonlinear Compton step the same amount for charges incoming energies $s\eta$ and $(1-s)\eta$. Charges with a higher initial energy will on average lose more momentum than those with a lower initial energy (illustrated in \figref{fig:NLC-momentum1}).
Contribution of the NLO process generates a skewness in the electron and positron lightfront momentum spectra, which can be further exacerbated by increasing the background field intensity or including NNLO processes of two photon emission $\gamma\to e^{+}e^{-}+\gamma'+\gamma''$, etc.

\subsection{Mono-energetic photon source}

In order to study the higher order contributions to the positron spectrum, we simulate a collision between an unpolarised photon beam and a high-intensity laser, such that positrons are produced by the nonlinear Breit-Wheeler process, and track the number of photons emitted by the positrons. (The electrons can emit an arbitrary number of photons.)
We then group them by emission count and examine how the skewness of the positron distributions changes as we include higher orders, as shown in \cref{fig:SeriesDiag1}. The skewness, $\mathcal{S}$, is defined as
    \begin{equation}
    \mathcal{S} = \frac{1}{N_{e^+}} \int_0^1 \!
        \left( f - \frac{1}{2} \right)^3 \frac{d N_{e^+}}{d f}
    \, d f
    \label{eq:SkewnessDef}
    \end{equation}
where $N_{e^+}$ is the number of positrons and $f = \epsilon / \omega$ is the ratio of the positron and photon energies. Focussing on experiment, simulation results will be presented in terms of energy rather than the lightfront momentum because emission is, to a good approximation, forward, with negligible transverse momentum gained from the laser ($\omega, \epsilon \gg m a_0$), meaning the lightfront momentum fraction $s$ is very close to the energy fraction $\epsilon / \omega$.
The relationship between moments of the energy distribution, including skewness, and radiation reaction (when all orders of emission are included) is discussed in \cite{niel.pre.2018}.

\begin{figure}[h!!]
\centering
$\underbrace{\raisebox{-.5\height}{\includegraphics[width=2.5cm]{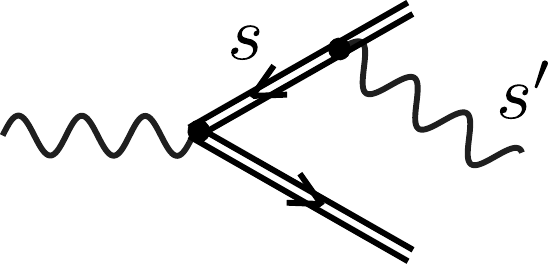}}}_{(2A)}$\hspace{0.5cm},\hspace{0.5cm}$\underbrace{\raisebox{-.5\height}{\includegraphics[width=2.5cm]{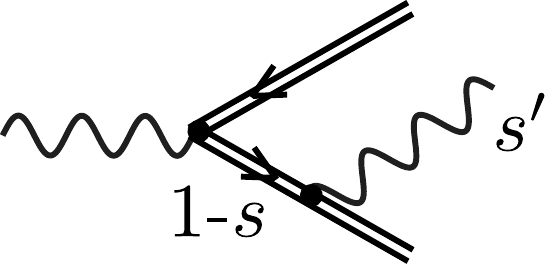}}}_{(2B)}$
\caption{
The two diagrams included in calculating the probability, $\tsf{P}^{(2)}$, of the $O(\alpha^{2})$ NLO process of photo-trident. The contribution of each diagram is  evaluated in the `long-pulse limit' where they are each factorised as an integral over sequential first-order building blocks,  and under the assumption that the probability is much smaller than unity}. \label{fig:d1}
\end{figure}

\begin{figure}[h!!]
\centering
$\underbrace{\raisebox{-.5\height}{\includegraphics[width=2.5cm]{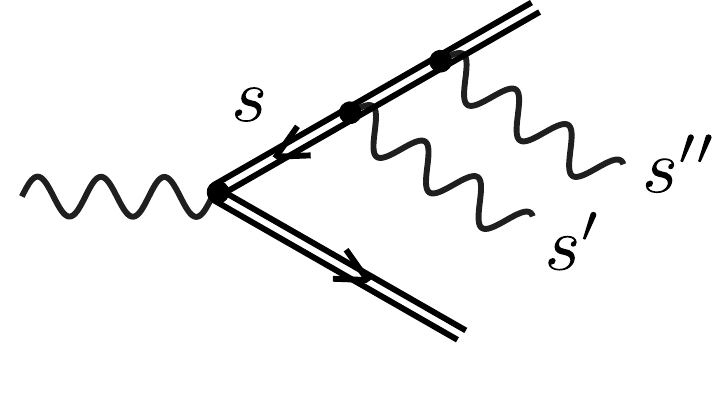}}}_{(3A)}$\hspace{0.5cm},\hspace{0.5cm}$\underbrace{\raisebox{-.5\height}{\includegraphics[width=2.5cm]{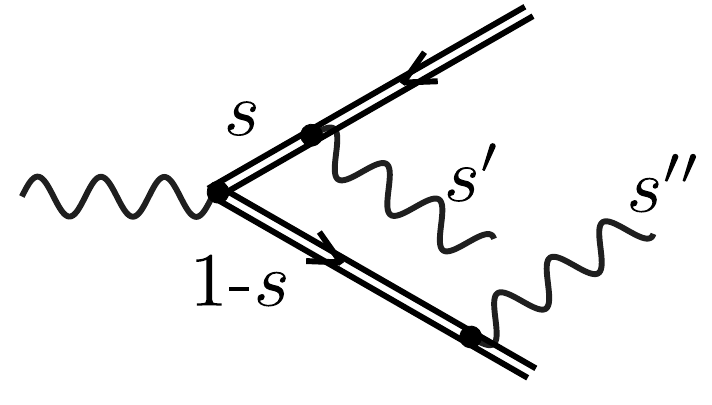}}}_{(3B)}$\hspace{0.5cm}, \\ \hspace{0.5cm}$\underbrace{\raisebox{-.5\height}{\includegraphics[width=2.5cm]{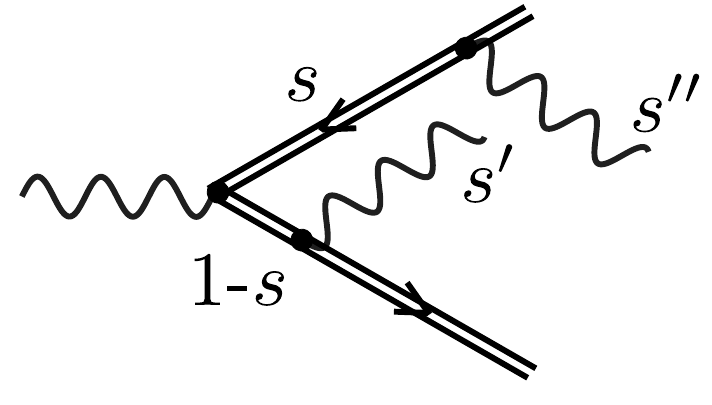}}}_{(3C)}$\hspace{0.5cm},\hspace{0.5cm}$\underbrace{\raisebox{-.5\height}{\includegraphics[width=2.5cm]{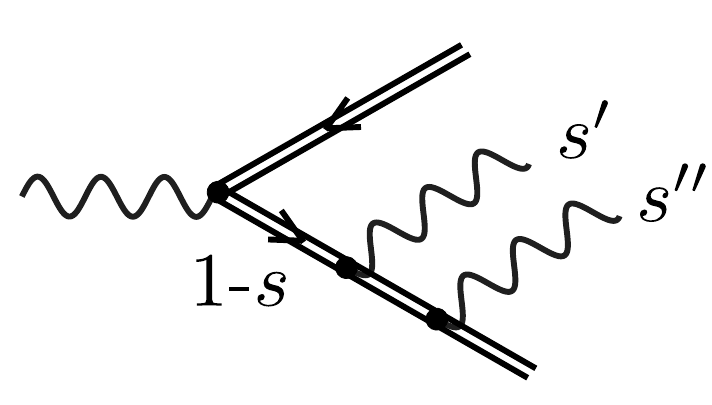}}}_{(3D)}$
\caption{
The four diagrams included in calculating the probability, $\tsf{P}^{(3)}$, of the $O(\alpha^{3})$ NNLO process of photo-trident. The contribution of each diagram is  evaluated in the `long-pulse limit' where intermediate propagators are on-shell and interference between diagrams is neglected.
} \label{fig:d2}
\end{figure}

\begin{figure}[h!!]
\includegraphics[width=8.2cm,draft=false]{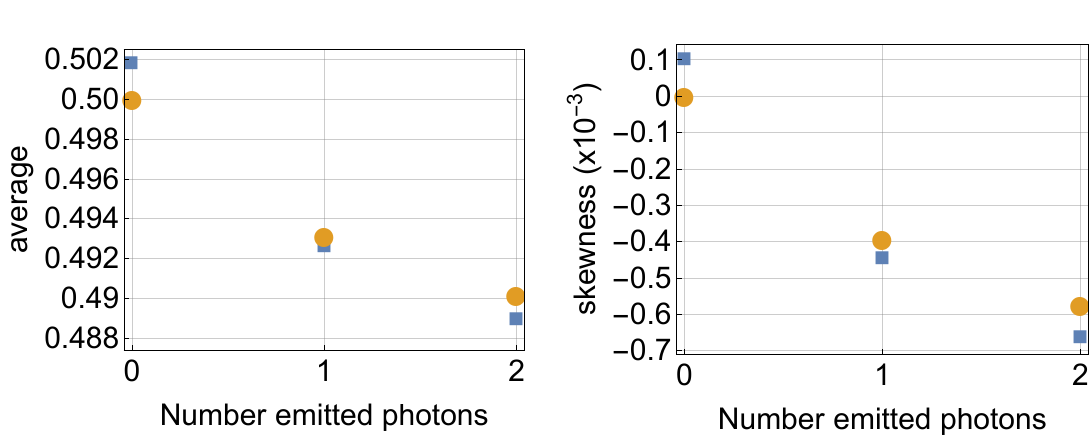}
    \caption{Comparison of the average lightfront momentum and skewness in the momentum spectrum of positrons produced by a photon of fixed energy parameter $\eta=0.2$, in an $N=8$ cycle sine-squared pulse with $a_{0}=5$.  (Blue) squares are simulation results and (orange) circles are from the direct calculation using the LCFA.}
    \label{fig:TheoryBench1}
\end{figure}

We can investigate how the skewness arises in the positron spectrum by direct evaluation of higher orders using first-order building blocks using the locally constant field approximation (LCFA) \cite{ritus.jslr.1985,Ilderton:2018nws,DiPiazza:2018bfu}. (A similar procedure has been employed for higher-order processes \cite{DiPiazza:2010mv,Torgrimsson:2021wcj,Torgrimsson:2021zob}.) These direct `theory' results can then be compared with the output of simulations. It is instructive to consider the NLO process, which comprises the two channels shown in \figref{fig:d1}. 
The lightfront momentum spectrum of each diagram can be written:
\bea
F^{\tsf{(2A)}}(s,s')  &=& \int^{\Phi}_{0}d\phi \frac{d\tsf{R}_{e}[\xi(\phi);\eta, s\eta]}{ds} \nn \\ 
&& \times \int_{\phi}^{\Phi}d\phi'\frac{d\tsf{R}_{\gamma}[\xi(\phi');s\eta,s'\eta]}{ds'}\nn \\
F^{\tsf{(2B)}}(s,s')  &=& \int^{\Phi}_{0}d\phi \frac{d\tsf{R}_{e}[\xi(\phi);\eta, (1-s)\eta]}{ds} \nn \\ 
&& \times \int_{\phi}^{\Phi}d\phi'\frac{d\tsf{R}_{\gamma}[\xi(\phi');(1-s)\eta,(1-s)s'\eta]}{ds'}\nn 
\eea
where $\tsf{R}_{e}$ and $\tsf{R}_{\gamma}$ are the probability per unit phase for NBW pair creation and NLC scattering respectively, and the arguments given are the intensity, energy parameter of incoming particle and energy parameter of produced particle (positron or photon) and the relation to the total probability at NLO is:
\[
\tsf{P}^{(2)} = \int_{0}^{1}ds \int_{0}^{1}ds' \left[F^{\tsf{(2A)}}(s,s')+F^{\tsf{(2B)}}(s,s')\right],
\]
assuming that the total probability is much smaller than unity. Defining $u$ to be the positron lightfront parameter after NLO scattering processes, we see that $u=s(1-s')$ in diagram $(\tsf{2A})$, whereas $u=s$ in $(\tsf{2B})$. Changing variables and integrating, we have:
\[
\frac{d\tsf{P}^{(2)}}{du} = \int_{u}^{1} F^{\tsf{(2A)}}\left(s,1-\frac{u}{s}\right)\frac{ds}{s} +\int_{0}^{1} F^{\tsf{(2B)}}\left(u,s'\right)ds'
\]
(the integration limits arise in the first integral due to the condition that ${0<s'<1}$). To calculate the NNLO, we follow a similar procedure, but this time there are four diagrams, shown in \figref{fig:d2}. 
The total positron spectrum formed by the LO, NLO and NNLO contribution is then:
\[
\frac{d\tsf{P}}{du} = \sum_{j=1}^{3}\frac{d\tsf{P}^{(j)}}{du}.
\]
Calculating moments $\av{u^{n}} = \int u^{n}(d\tsf{P}/du) du$, a theory estimate of the skewness is $\tsf{S}=\av{(u-1/2)^3}/\av{u^{0}}$. (Note the theory definition of skewness uses lightfront momentum variables in contrast to \eqnref{eq:SkewnessDef}.)  To benchmark with simulation, we consider a sine-squared plane-wave pulse with intensity parameter:
\[
a_{0}(\varphi) = a_{0} \sin^{2}\left(\frac{\varphi}{2N}\right)\cos\varphi,
\]
for $0<\varphi<2\pi N$ and $a_{0}(\varphi) = 0$ otherwise where $N$ is the number of laser cycles. Then choosing $N=8$, $a_{0}=5$ and $\eta=0.2$ (corresponding to around $16\,\trm{GeV}$ electrons for an almost head-on collision with the laser), and using spin-averaged LO building blocks, we find agreement to within around 10\% in the average lightfront parameter and skewness, as shown in \figref{fig:TheoryBench1}. In this benchmark, we neglect the propagating photon's polarisation, which is known to lead to around $\sim 10\%$ error in the LCFA \cite{King:2013zw} and a similar level in the LMA \cite{Tang:2022ixj}. However, the photon polarisation is not expected to affect the kinematic signatures investigated here. (Furthermore, photon polarisation \emph{is} included in the rest of the simulation results.)

\begin{figure}
    \centering
    \includegraphics[width=1\linewidth]{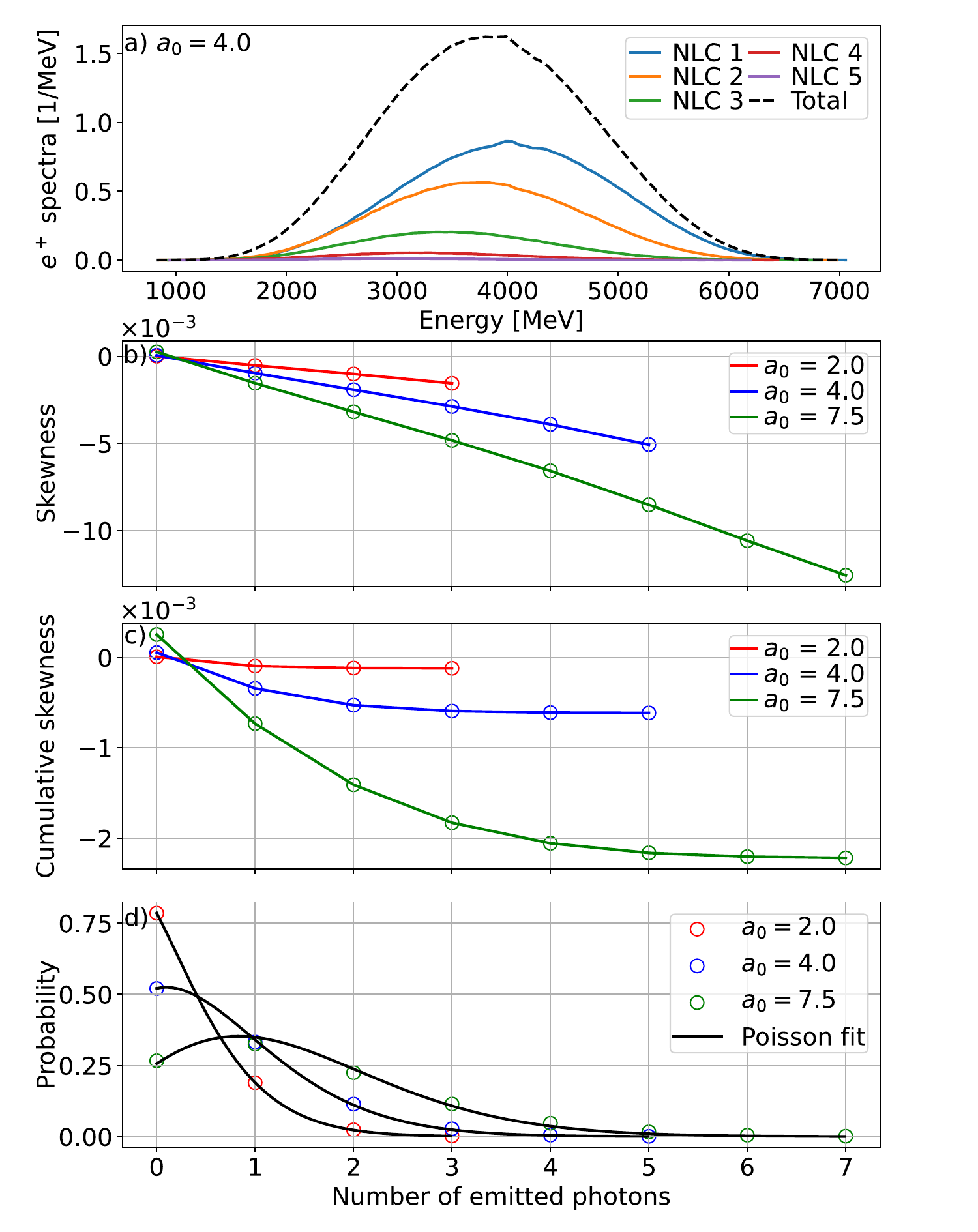}
    \caption{a) Positron energy spectra for a laser with $a_0=4.0$, separated by the number of nonlinear Compton events.
    The dashed line indicates the sum of all contributions.
    b) Skewness of $n$-fold nonlinear Compton scattering for laser amplitudes $a_0=2.0$, $4.0$ and $7.5$.
    c) Cumulative skewness of the distributions in panel a), as defined by \eqnref{eq:SkewnessDef}, for laser amplitudes $a_0=2.0$, $4.0$ and $7.5$.
    d) The fraction of positrons that emit the specific number of photons only (i.e. the multiplicity), in collisions at the given $a_0$.
    (Solid lines are fitted Poisson distributions.)}
    \label{fig:positron_photon}
\end{figure}

Following this benchmarking we return to simulation, which allows us to investigate even higher orders  and emission. The simulation is of electron-positron creation when a mono-energetic, unpolarised 8 GeV photon beam collides head on with a laser of Gaussian temporal profile with full width at half maximum duration $\tau= 30$ fs and $\lambda= 0.8$ µm.  The difference between the positron spectra for different photon emission counts (multiplicity) for $a_{0}=4$ is shown in \figrefa{fig:positron_photon}. (For this comparison we choose an $a_0$ that is not too large to restrict the number of secondary emissions and thereby control the onset of higher order contributions.) The positrons that emit zero photons via nonlinear Compton scattering (NLC 0, solid blue line) are symmetric around 4000 MeV, which corresponds to $s = 1/2$. However, for positrons that emit one photon (NLC 1) or more, the distributions become increasingly negatively skewed, and the peak moves down in energy. If we calculate the total contribution from all emissions we get the dashed line, which is also negatively skewed around $s = 1/2$. For higher orders, i.e. the emission of more photons, the average total positron recoil becomes larger and hence increases the skewness in the distribution.

The effect of the photon multiplicity on the skewness of the positron distribution is shown in \figref{fig:positron_photon}b) and c) for various values of $a_0$. As one may expect, the distributions for zero photon emission are all nearly symmetric; the asymmetry grows as more photons are emitted. However, the skewness grows faster with greater $a_0$ because the positron $\chi$ parameter is larger and therefore individual photons are more energetic.
As more higher orders are included, the total skewness, summed over all contributions, saturates at a negative value.

In \figref{fig:positron_photon}\,d) the probability to emit a given number of photons is plotted for the same $a_0$ as in \figrefb{fig:positron_photon}. With increasing $a_0$ the most probable multiplicity is shifted towards higher orders. Since higher-order nonlinear Compton scattering is included in numerical simulation by repeated Monte Carlo sampling of single nonlinear Compton scattering, the distributions can be seen to agree well with classical Poisson statistics ($\chi \ll 1$). If the individual events occur independently of each other, then the number of events are distributed accordingly: 
\bea 
\tsf{P}(n\gamma) = \frac{\mbox{e}^{-\mu}\mu^{n}}{n!},
\eea
where the average number of photons $\mu = \int W dt$ for emission rate $W$. We calculate a fit to the simulation data using the average number of emissions as an initial guess of $\mu$ to generate the curves in \figref{fig:positron_photon}\,c). The agreement is good with a slight deviation due to the quantum effect of recoil, which should cause the distribution to be non-Poissonian due to memory effects. Further deviations due to dressed loops and virtual pathways could also be tested in photon-laser collisions.

\subsection{Photon source with finite bandwidth}

All of this suggests that the skewness of the positron distribution may be used as a signal of higher order contributions. However if the initial photon beam is not monoenergetic, then skewness will arise even at leading order. Therefore, let us consider the case where the photon beam has a Gaussian energy spectrum and bandwidth $0 \leq \delta \omega / \omega < 10\%$ root-mean-squared (RMS), keeping $a_0$ fixed at $a_{0}=4$ and retaining the other parameters as before.
The corresponding full-width-at-half-maximum (FWHM) $(\Delta \omega / \omega)_\text{fwhm} = 2.35 (\delta \omega / \omega)$.

\begin{figure}
    \centering
    \includegraphics[width=1\linewidth]{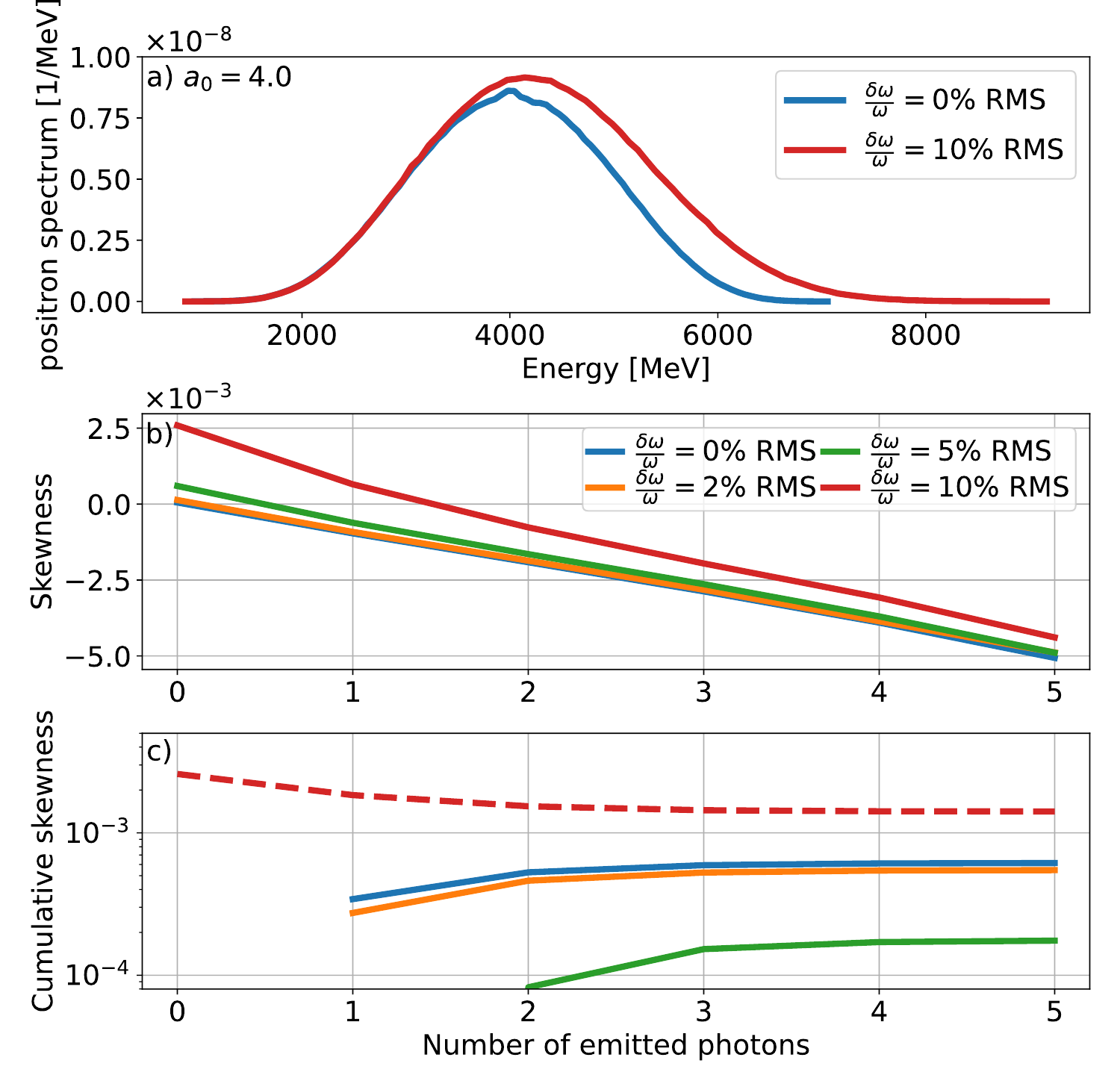}
    \caption{a) Energy spectra of positrons created by a photon beam of energy 8 GeV and bandwidth of 0\% or 10\% (RMS) in a laser with amplitude $a_0 = 4.0$, at lowest order (i.e. zero emitted photons). b) The skewness of the positron spectra for higher order contributions, for various bandwidths.
    c) The absolute value of cumulative skewness of the positron spectra for higher order contributions, for various bandwidths. Solid (dashed) lines indicate that the sign is negative (positive).}
    \label{fig:skewness_variance}
\end{figure}
In \figref{fig:skewness_variance}a) the positron spectrum at lowest order (NLC 0) is shown, given that the photon beam has a bandwidth 0\% RMS (solid blue) and 10\% RMS (dashed purple). At LO and zero bandwidth, the spectrum is the same as in \figref{fig:positron_photon}a), but for the 10\% bandwidth case, the spectrum is clearly positively skewed. This arises because higher energy photons are more likely to produce pairs, so that the high energy tail of the positron spectrum is enhanced.

The skewness of the different positron distributions at different multiplicities and bandwidths is shown in \cref{fig:skewness_variance}b), with the cumulative skewness in \cref{fig:skewness_variance}c). Here we see that for both 0\% and 2\% RMS (blue and orange lines) that the lowest order has a skewness close to zero. As the bandwidth increases above 2\%, so does the skewness at LO. However, for higher numbers of emitted photons, the skewness becomes increasingly negative, since recoil becomes more important, as demonstrated in \cref{fig:positron_photon}b). Therefore in order to observe higher order contributions, we need the bandwidth-induced skewness at LO to be smaller (in absolute terms) than the recoil-induced skewness in the absence of an initial bandwidth. For example, at NLO the recoil-induced skewness for a 0\% bandwidth is 0.08. In order for the bandwidth-induced skewness at LO to be smaller than this, the bandwidth must be lower than 5\% (where the skewness is 0.07). One therefore requires a quasi-monoenergetic photon source, such as linear Compton scattering of a high-quality electron beam \cite{Seipt_2025}, to observe this signal of higher order effects.

\section{Higher Order Nonlinear Processes}

\subsection{Higher order nonlinear Compton scattering}
\label{sec:HONLC}

To highlight the signals of higher-order nonlinear Compton scattering, we begin by recapping some of the signals of \emph{single} nonlinear Compton scattering. Again using the Ptarmigan \cite{blackburn.pop.2023} simulation code, but this time for the collision of an $8\,\trm{GeV}$ \emph{electron} beam with a linearly polarised optical laser pulse having a temporal Gaussian envelope with FHWM duration $\tau=30\,\trm{fs}$, a scan over $a_0$ is performed in the range $0.1 < a_0 < 10$ by varying the laser spot size $w_0$, while keeping the laser pulse energy fixed at a value of 6.0~J ($\mathcal{E} \propto a_0^2 w_0^2 = \text{const}$).

\begin{figure}[!!h]
        \centering
        \includegraphics[width=1\linewidth]{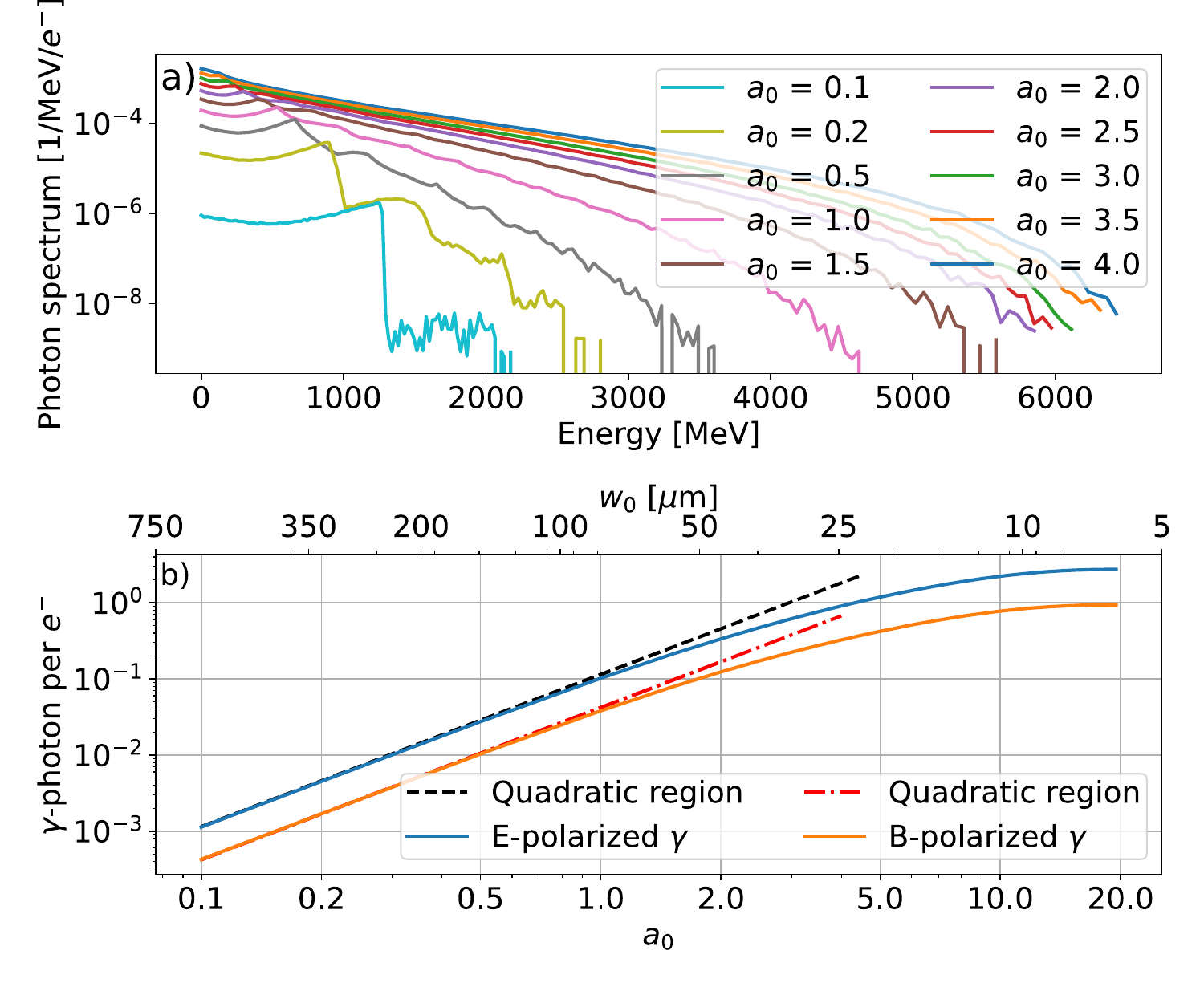}
        \caption{a) Photon energy spectrum for increasing $a_0$. b) The total number of photons per electron as a function of $a_0$ (lower axis) and $w_0$ (top axis). For $a_0 \geq 25$ the waist $w_0$ is smaller than the electron-beam size $\rho = 3~\mu$m and the photon yield begins to fall.}
        \label{fig:PhotonSpectra}
    \end{figure}

In the limit of low $a_0$, Compton scattering is linear, whereas for larger $a_0$ nonlinear effects can be seen: the Compton edge is pushed to lower energy when $a_0$ increases (see \figrefa{fig:PhotonSpectra}). The width of the emitted photon spectrum increases with $a_0$ because the probability to absorb increasingly many photons grows with laser intensity. \Cref{fig:PhotonSpectra}b) shows the total number of photons produced per electron, resolved into the two polarisation components, as a function of $a_0$. The collision produces photons that are mostly E-polarised, where E-polarised means that the photons, emitted head-on to the laser pulse, are polarised parallel to its electric field, and B-polarised that they are polarised parallel to the magnetic field. The dashed lines indicate the transition from the low-intensity, $a_{0}\ll 1$, perturbative regime in which Compton scattering is linear and the high-intensity, $a_{0} \gtrsim O(1)$, all-order regime, in which there is a `turning of the curve' of the yield of photons away from the simple $\sim a_{0}^{2}$ scaling. This dependence change has been  measurable in experiment at low strong-field parameter $\chi \ll 1$ \cite{Yan:2017dwf} and should be measurable for the parameters we consider here because of the high numbers of photons produced and because the laser spot size is larger than the electron beam. The laser spot size decreases with $a_0$ and for $a_0\gtrsim25.0$ it becomes narrower than the electron beam waist, $\rho =3$ $\mu$m. If the intensity is increased above this, the change in yield will be due to both the different scaling with $a_{0}$ and the  reduced overlap of the beams.

\begin{figure}
    \centering
    \includegraphics[width=1\linewidth]{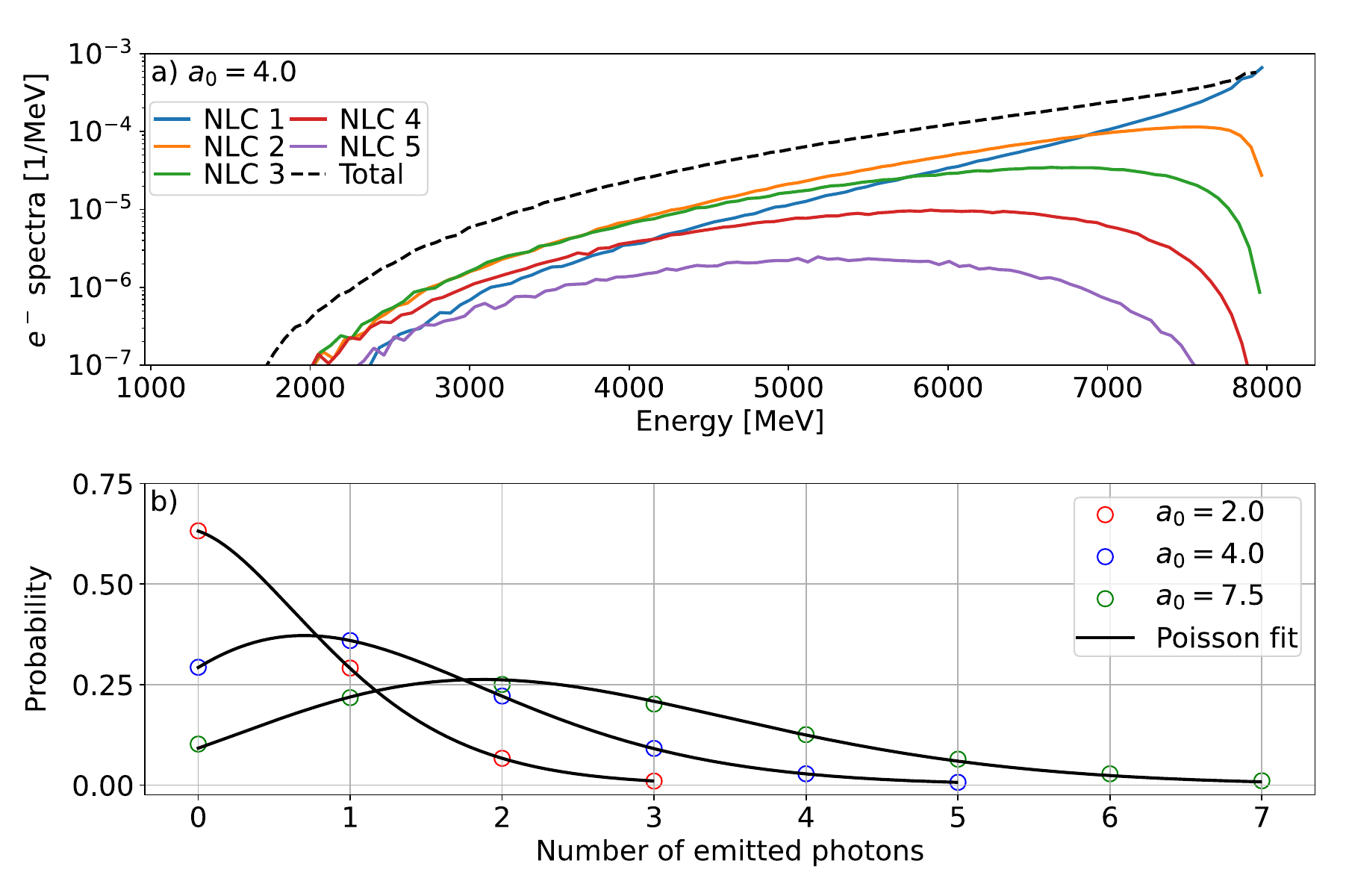}
    \caption{a) Post-collision electron spectra, separated by the number of emitted photons, at $a_0 = 4$. b) The fraction of electrons that emit the specific number of photons $N$ in collisions at the given $a_0$.}
    \label{fig:multiplicity}
\end{figure}

For \emph{higher}-order nonlinear Compton scattering, we focus on deviations from the single nonlinear Compton spectra in \figref{fig:PhotonSpectra}\,a). By enabling multiplicity tracking in the code, we can identify which electrons have radiated a given number of times.
\figref{fig:multiplicity}\,a) shows the electron spectrum split into components with different multiplicities. As described in the introduction, for `long pulses' (number of laser cycles $\gg 1$) and $a_{0}^{2} \gg 1$, $n$th order nonlinear Compton scattering from a single electron is believed to be well-estimated by $n$ sequential first-order emissions, providing the dominant contribution to the probability at $O\left(\alpha^{n}\right)$~\cite{fedotov.pr.2023}. On the other hand, the inclusion of loops  is important for unitarity and affects the summation of diagrams. Higher-order strong-field QED can be tested by measuring the electron beam energy distribution following nonlinear Compton showers. In \figref{fig:multiplicity}\,a), we see that the electrons that lose most energy have done so by undergoing higher-order nonlinear Compton scattering. This partially reverses the usual perturbative hierarchy. The contribution from NLC 1 (emission of one photon only) falls below than that of NLC 2 (emission of two photons) for energies smaller than 6.8~GeV. As the energy is reduced further, higher order processes become increasingly important: NLC 1 falls below NLC 3 at 5.7 GeV and then NLC 4 at 4.0 GeV. Precision measurement of the spectrum in this energy range could then identify the dominance of these contributions. 

The multiplicity is plotted as a function of $a_0$ in \figrefb{fig:multiplicity}. With increasing $a_0$ the most probable multiplicity is shifted towards higher values. Since higher-order nonlinear Compton scattering is included in numerical simulation by repeated Monte Carlo sampling of single nonlinear Compton scattering, the distributions can be seen to agree well with classical Poisson statistics ($\chi \ll 1$), with a slight deviation due to the quantum effect of recoil. Further deviations due to dressed loops could also be tested in electron-laser collisions.

\subsection{Higher order nonlinear trident}
\begin{figure}[h!!]
\includegraphics[width=6.5cm]{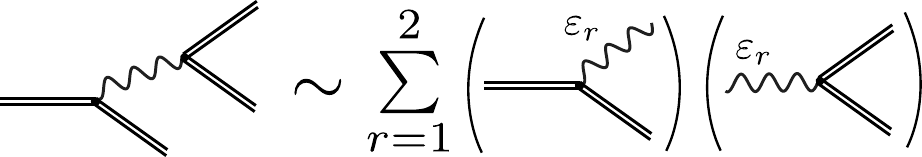}
\caption{In the `long-pulse limit', nonlinear trident process is well approximated as nonlinear Compton scattering of a real polarised photon and its subsequent conversion to an electron-positron pair via the nonlinear Breit-Wheeler process (the sum over $r$ implies a sum over photon polarisations).}\label{fig:NLtridentDiagram}
\end{figure}

A process with a clear signature of higher-order strong-field QED is nonlinear trident. When $a_{0}\Phi\gg 1$ it is expected that nonlinear trident can be well-approximated as a sequential product of the nonlinear Compton scattering of a polarised photon and that photon's conversion to an electron-positron pair via nonlinear Breit-Wheeler pair-creation \cite{fedotov.pr.2023} (depicted in \figref{fig:NLtridentDiagram}). Since the nonlinear Breit-Wheeler step is usually less probable than the Compton step, the probability of the Breit-Wheeler step normally determines the probability of nonlinear trident.

The yield of positrons produced when the electron beam collides with the optical laser is a clear test of the nonlinear interaction. When $a_{0} \ll 1$, the yield of pairs depends on $a_{0}^{2n_{\ast}}$, where $n_{\ast}$ is the threshold number of laser photons required to produce an electron-positron pair and the interaction is \emph{multi-photon}, that is, perturbative, but with the leading order being highly nonlinear in the laser field. The threshold number of photons is, $n_{\ast}=\lceil 2(1+a_{0}^{2})/\eta_{\gamma} \rceil$, where $\eta_{\gamma}$ is the energy parameter of the photon that produces the pair. Since $\eta_{\gamma} \leq \eta$, and $\eta \sim O(0.1)$ for the all-optical parameters considered here, clearly $n_{\ast} \gg 1$. As $a_{0}$ increases such that $a_{0}\gtrsim 1$, the yield of positrons deviates from the power-law scaling. This `turning of the curve' is a strong signal of the all-order interaction between the laser and the electron-positron pair \cite{abramowicz.epjst.2021}.

\begin{figure}[h!!]
        \centering
\includegraphics[width=1\linewidth]{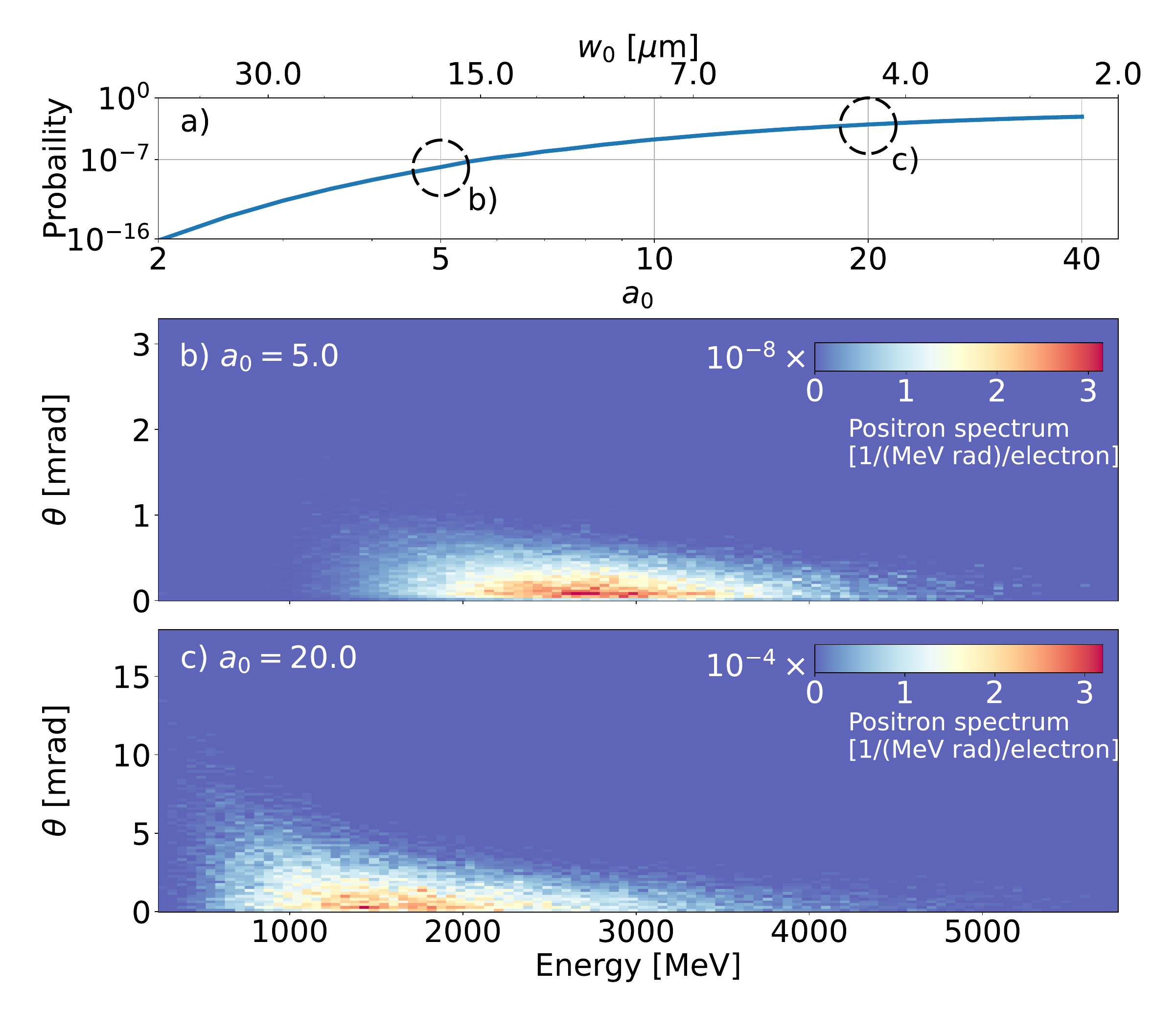}
        \caption{a) Probability of creating a positron (i.e. positron yield per electron) as a function of $a_0$. If $a_0\leq 1$, the probability $< 10^{-30}$.
            The laser focal spot size $w_0$ required to achieve the given $a_0$ is shown on the upper horizontal axis. Positron spectrum for (b) $a_0 = 5.0$ and (c) $a_0= 20.0$, over energy and angle.}
        \label{fig:PositronProb}
    \end{figure}

Again considering the collision of an electron beam of initial energy $8\,\trm{GeV}$ with a laser pulse with Gaussian spatial and temporal envelopes (FWHM duration $\tau=30\,\trm{fs}$, fixed energy 6.0~J, as in \cref{sec:HONLC}), we plot the probability for nonlinear trident as a function of the intensity parameter $a_{0}$, calculated by Ptarmigan, in \figref{fig:PositronProb}. The multi-photon regime is not accessible because the positron yield is too small to be detected; however, the turning of the curve that provides a signal of the all-order interaction, \emph{is} accessible. Since it is determined by $a_{0}$, which is a \emph{classical} parameter, the all-order interaction is `\emph{non-perturbativity at small coupling}' is classical, even though it is a prediction of quantum theory. However, if pairs are created when $a_{0} \gg 1$ but $\chi \ll 1$, the process is well-described as being in the quantum `tunneling' regime. Being controlled by the quantum parameter $\chi$ and scaling as $\sim \exp[-16/3\chi]$ \cite{Ritus:1972nf} this process is entirely quantum in nature. 
Literature results show \cite{King:2024ffy} that if nonlinear trident pairs can be observed in experiment when $a_{0} \approx 5$, for electron beam energies similar to considered here, tunneling pair-creation and its non-analytic dependency on the coupling can be measured. From the results in \figref{fig:PositronProb}, we see that for parameters considered here at $a_{0}=5$, approximately one pair is generated for every $\sim 10^{8}$ electrons colliding with the optical beam.
The positron energy spectra at $a_{0}=5$ and $a_{0}=20$ are shown as insets to \figref{fig:PositronProb}. At $a_{0}=5$, the angular spread of positrons is rather narrow, but at $a_{0}=20$, the low-energy tail is clearly spread out more than the rest of the distribution.
As seen for nonlinear Compton scattering in \figref{fig:multiplicity}, the shift towards low energy is a signal of higher-order processes: the energy loss is greater due to more photons being emitted. There is also an angular spread which is larger for lower energies, which may be explained by the stochastic transverse spreading of electrons and positrons discussed in \cite{PhysRevLett.112.164801}: here a greater angular spread indicates that a greater number of photons has been emitted. In fact, higher order events are inevitable in nonlinear trident, where the lowest order contribution is to create a pair and emit no photons. This has to be very unlikely since $P(\text{no photons}\,|\,\text{pair created}) = 1- P_\gamma$ (where $P_\gamma$ is the probability to emit a photon) and for any $a_0$ large enough to create a reasonable number of pairs, $P_\gamma\sim 1$. Again, to see evidence of this higher-order signal, precision measurement of the energy-angle spectrum is required. One can draw a similar conclusion from recent measurements of nonlinear trident in electron-crystal collisions where a particular part of the positron and pair energy spectrum has hinted at deviations from the expected two-step result \cite{Nielsen:2023icv}.

\section{Conclusion}

The strong-field QED science goals of current and emerging multi-PW laser facilities include the important aim of measuring and characterising for the first time leading-order nonlinear Breit-Wheeler pair-creation and nonlinear Compton scattering at strong-field parameters $\chi \sim O(1)$. Although experimental control of these processes must still be developed, the theory and phenomenology of these processes have been studied in depth and they are believed to be well understood. On the other hand at very high strong-field parameter, $\chi\sim O(1000)$, the Ritus-Narozhny (RN) conjecture \cite{Fedotov:2016afw} predicts that strong-field QED becomes completely non-perturbative (in charge-background and charge-radiation-field interaction) and there are currently no clear predictions from theory in this regime. 
In this paper our interest has been in highlighting some experimental signals in the transition between these two regimes. We referred to these as \emph{nonlinear showers} (in contrast \emph{nonlinear avalanches} where the reacceleration of intermediate particles by the laser field is important \cite{bell.prl.2008,bulanov.pre.2011,ridgers.prl.2012,King:2013zw,Mironov:2014xba,meuren.prd.2016,grismayer.pre.2017,gonoskov.prx.2017,Mironov:2025nhk,mercuribaron.prx.2025}). It is believed that the current calculational framework is accurate in these regimes, but if the RN conjecture is correct, at some point it will break down. This motivated our interested in studying signals of higher order strong-field QED processes, which must be confronted by experiment.

Specifically, we have focussed on nonlinear photo-trident and showed that the asymmetry (skewness) in the lightfront momentum spectrum of positrons and electrons can act as a clear signal of the second-order and possibly higher-order processes. A key requirement for observing this signal is that the skewness in the leading-order process (nonlinear Breit-Wheeler pair creation), induced by the bandwidth of the photon source, must not be much larger than the the total skewness induced by higher-order contributions. For nonlinear Compton showers, we showed that the contribution from higher-order processes to the low-energy tail of scattered electrons can already dominate for $a_{0} \sim O(1)$. Finally, for nonlinear trident an asymmetry in the electron-positron lightfront momentum spectrum at $a_{0}\sim O(10)$ can, just like for phototrident, be used as a signal of higher-order nonlinear shower processes. Therefore, efforts to produce a higher quality electron beam and detection methodology, whilst having a good understanding of the magnitude of experimental and theory errors \cite{blackburn2025theoryuncertaintiespredictionsstrongfield}, can give important insights to the role of higher orders of strong-field QED processes.

\section*{Acknowledgments}
TB, BK and MS acknowledge the organisers of PHFS2023, the 79th Fujihara seminar, where some of these ideas were first developed. BK acknowledges support from The Leverhulme Trust, Grant No. RPG-2023-285.

\appendix

\bibliography{references}

\end{document}